\documentclass[11pt]{article}
\usepackage{latexsym}
\usepackage{amsmath} 
\usepackage{amscd}     \usepackage{amsxtra}
\usepackage{upref}     \usepackage{amsthm}
\usepackage{amssymb}   
\usepackage{amsfonts}
\begin{document}

\title{{\bf Parallel Objects and Field Equations}}

\author{{\bf Stoil Donev}
\\ Institute for Nuclear Research and Nuclear
Energy, Bulg.Acad.Sci.,\\ 1784 Sofia, blvd.Tzarigradsko chaussee 72,
Bulgaria\\ e-mail: sdonev@inrne.bas.bg\\
{\bf Maria Tashkova}\\Institute of Organic Chemistry with Center of
Phytochemistry,\\ Bulg. Acad. Sci., 1113 Sofia, Acad.
G.Bonchev Str.  9, Bulgaria}

\date{}
\maketitle

\begin{abstract}
This paper considers a generalization of the existing
concept of parallel (with respect to a given connection) geometric objects
and its possible usage as a suggesting rule in searching for adequate field
equations in theoretical physics. The generalization tries to represent
mathematically the two-sided nature of the physical objects, the {\it
change} and the {\it conservation}. The physical objects are presented
mathematically by sections $\Psi$ of vector bundles, the admissible changes
$D\Psi$ are described as a rsult of the action of appropriate
differential operators $D$ on these sections, and the conservation
propertieis are accounted for by the requirement that suitable projections of
$D\Psi$ on $\Psi$ and on other appropriate sections must be zero. It
is shown that the most important equations of theoretical physics obey this
rule.  Extended forms of Maxwell and Yang-Mills equations are also
considered.

\end{abstract}



\section{Introduction}
When we think about physical objects, e.g. classical particles,
solid bodies, elementary particles, etc., we always keep in mind that,
although we consider them as free, they can not in principle be absolutely
free, otherwise they would be undectable.  What is really understood under
"free object" is, that some {\it definite properties (e.g. mass, velocity) of
the object under consideration do not change in time under the influence of
the existing enviorenment}.  The availability of such time-stable features of
any physical object guarantees its {\it identification} during its existence
in time.  Without such an availiability of constant in time properties, which
are due to the object's resistense abilities, we could not speak about
objects and knowledge at all.  So, a classical mass particle in external
gravitational field is free with respect to its mass, and it is not free with
respect to its behaviour as a whole, because in classical mechanics formalism
its mass does not change during the influence of the external field on its
accelareted way of motion.

In trying to formalize these views we have to give some initial explicit
formulations of some most basic features (properties) of what we call
physical object, which features would  lead us to a, more or less, adequate
theoretical notion of our intuitive notion of a physical object.
Anyawy, the following properties of the theoretical concept "physical object"
we consider as necessary:

      1. It can be created.

      2. It can be destroyed.

      3. It occupies finite 3-volumes at any moment of its existence, so it has structure.

      4. It has a definite stability to withstand some external disturbances.

      5. It has definite conservation properties.

      6. It necessarily carries energy-momentum, and, possibly,
other measurable (conservative or nonconservative) physical quantities.

      7. It exists in an appropriate environment (called usually vacuum), which
provides all necessary existence needs.

      8. It can be detected by the rest of the world through allowed
energy-momentum exchanges with the "rest of the world".

      9. It  may combine with other appropriate objects to form new
objects of higher level structure.

      10. Its death gives necessarily birth to new objects following definite
rules of conservation.

\vskip 0.3cm
Clearly, together with the purely qualitative features a physical
object carry important physical properties which can be described
quantitatively by corresponding quantities, and any interaction between
two physical objects is, in fact, an exchange of such quantities provided
both objects carry them. Hence, the more universal is a physical quantity the
more usuful for us it is, and this moment determines the exclusively
important role of energy-momentum, which modern physics considers as the most
universal one, i.e. no physical objects are known that carry no
energy-momentum.

If we can identify a given physical object, represented locally in space-time
by the mathematical object $\Psi$, at different moments of its existence,
this means that the changes $D\Psi$ of its time-changing properties
vanish when are {\it "projected" upon the same object}, making use of other
appropriate objects $Q$.  From formal point of view this means that some
mathematical expression of the kind $\mathbb{F}(\Psi,D\Psi;Q)=0$,
specifying what and how changes, and specifying also what is projected and
how it is projected, should exist.  Hence, specifying differentially some
conservation properties of the system under consideration, we obtain
equations of motion being consistent with these conservation properties.  We
recall that this idea has been used firstly by Newton in his momentum
conservation equation $\dot\mathbf{p}=\mathbf{F}$, which is the restriction
of the partial differential system $\nabla_{\mathbf{p}}\mathbf{p}=\mathbf{F}$
on some trajectory. This Newton's system of equations just says that there
are physical objects in Nuture which admit the "point-like" approximation,
and which can exchange energy-momentum with "the rest of the world" but keep
unchanged some other intrinsic properties which allows their identification
in time.

These two aspects of any physical object (or a system of objects) - {\it
change} and {\it conservation}, - have been very successfully unified and
presented as a working tool (computational prescription) by the {\it
variational procedure} (Lagrange-Euler-Hamilton action priciple). The central
idea of this approach is that if something happens, i.e. some real process
develops, in some 4-dimensional region in Nature, there is an {\it
optimization} quantity characterizing its optimal way of development. This
integral quantity has been  called {\it action}, and its local representative
is usually called {\it lagrangean} (or lagrangean density). If the lagrangean
is known the procedure works perfectly in almost all theoretical physics, and
gives explicit "equations of motion" and local "conserved quantities". The
need of such a powerful tool is out of doubt, especially in microphysics
where the system studied changes considerably during observation, and
moreover, we have to get knowledge of it in a very {\it indirect} way.
However, this approach has a formal nature, it does NOT prescribe the
lagrangeans, moreover, many lagrangeans give the {\it same} equations of
motion and integral conserved quantities. One needs initial knowledge of the
system (like symmetry and stability properties, dynamical behaviour features,
etc.) in order to guess the corresponding lagrangean. In field theory this
situation frequenly leads to studies of "model lagrangeans": scalar field,
vector field, spinor field, etc., and to separatation of "free field" terms
from "interaction" terms in a lagrangean. The free field terms are
meant to give the system's {\it intrinsic} dynamics, and the interaction
terms describe some external influences. While in macrophysics, as a rule,
the external influences are such that they do NOT destroy the system, in
microphysics a full restructuring is allowd: the old ingredients of the
system may fully transform to new ones, (e.g.  the electron-positron
annihilation) provided the energy-momentum conservation holds. The essential
point is that {\it whatever the interaction is, it always results in
appearing of relatively stable objects, carrying energy-momentum and some
other particular physically measurable quantities}.  This conclusion
emphasizes once again the importance of having an adequate notion of what is
called a physical object, and of its mathematical representation.

From the point of view of spatial extension the physical objects may be {\it
point-like, finite} and {\it infinite}, but realistic seems to be just the
second option (recall p.3 above), although the classical approximations of
point-like (i.e.  structureless) objects and the infinite plane waves have
served as good approximations wherever they have been
uncontradictionally introduced and used.  However, modern science requires a
better adequacy between the real objects and the corresponding mathematical
model objects.  So, the mathematical model objects $\Psi$ must necesserilly
be spatially finite, and even temporally finite if the physical object
considered has by its intrinsic nature finite life-time.  This most probably
means that $\Psi$ must satisfy nonlinear partial differential equation(s),
which should define in a consistent way the {\it admissible} changes and the
{\it conservation} properties of the object under consideration. Hence,
talking about physical objects we mean a spatially finite entities which have
a well established balance between change and conservation, and this balance
is kept by a permanent and strictly fixed interaction with the enviorenment.

The conservation properties of an object manifest themselves through
corresponding symmetry properties, and these physical symmetry properties
appear as mathematical symmetries of the corresponding equations
$\mathbb{F}(\Psi,D\Psi;Q)=0$ in the theory. Usually, responsible for
these symmetries are some new (additional) mathematical objects defining the
explicit form of the equation(s), e.g. the Minkowski pseudometric tensor
$\eta$ in the relativistc mechanics and relativistic field theory, the
symplectic 2-form $\omega$ in the Hamilton mechanics, etc.  Knowing such
symmetries we are able to find new solutions from the available ones, and in
some cases to describe even the whole set of solutions.  That's why the Lie
derivative operator (together with its generalizations and prolongations
[1]) and the integrability conditions for the
corresponding equation(s) $\mathbb{F}(\Psi,D\Psi;Q)=0$ play a very
essential and hardly overestimated role in theoretical physics. Of course,
before to start searching the symmetries of an equation, or of a mathematical
object $\Psi$ which is considered as a model of some physical object, we must
have done the preliminary work of specifying the mathematical nature of
$\Psi$, and this information may come only from an inital data analysis of
appropriately set and carried out experiments.

The mathematical concept of symmetry has many faces and admits various
formulations and generalizations. The simplest case is a symmetry of a
real valued function $f: M\rightarrow \mathbb{R}$, where $M$ is a
manifold, with respect to a map $\varphi: M\rightarrow M$: $\varphi$ is
a symmetry (or a symmetry transformation) of $f$ if $f(\varphi(x))=f(x), x\in
M$. If $\varphi_t, t\in(0,1)\subset\mathbb{R}$ is 1-parameter group of
diffeomorphisms of $M$, then the symmetry $f(\varphi_{t}(x))=f(x)$ may be
locally expressed through the Lie derivative $L_X(f)=0$, where the vector
field $X$ on $M$ generates $\varphi_t$. If $T$ is an arbitrary tensor field
on $M$ then the Lie derivative is naturally extended to act on $T$ and we
call $T$ symmetric, or {\it invariant} with respect to $X$, or with respect
to the corresponding (local, in general) 1-parameter group of diffeomorphisms
of $M$, if $L_{X}T=0$. In this way the Lie derivative represents an
universal tool to search symmetries of tensor fields on $M$ with respect to
the diffeomorphisms of $M$. Unfortunately, this universality of $L_X$ does
not naturally extend to sections of arbitrary vector bundles on $M$, where we
need additional structures in order to introduce some notion of symmetry or
invariance.

We may find a suggestion how to approach this problem by slightly changing
the point of view, namely, to look for those tensor fields $T$ on $M$ which
satisfy the equation $L_{X}T=0$, where the vector field $X$ is given. We
obtain in this way a system of differential equations for $T$, i.e.  we
search for the kernel $Ker(L_{X})$ of the differential operator $L_{X}$, and
call the solutions {\it symmetric}, or {\it invariant}, with rerspect to $X$.
So, we may generalize the situation to any (physically sensible) differential
operator $D: \Psi(x)\rightarrow (D\Psi)(x)$, where
$\Psi(x)$ are the fields of interest, (sections of appropriate vector
bundles), and to call the solutions of $D(\Psi)=0$ {\it symmetric,
(invariant)}, with respect to $D$.  As a rule, in such cases the
solutions carry appropiate names, for example, if $D=\nabla$, where
$\nabla$ is a linear connection in a vector bundle [2], then a section
$\sigma$ of this bundle is called {\it parallel} with respect to $\nabla$ if
$\nabla(\sigma)=0$.

We'd like to note that, in general, the symmetry of an object is {\it always}
with respect to something (group of transformations, differential operator,
sections of some vector bundles, etc.) preliminary fixed. And if the symmetry
we are looking for will be given some physical interpretation, the
preliminary work needed to fix the symmetry operator should be done by
theoretical physics.

Following the above stated views we are going to consider in this paper a
more generl view  on the geometrical concept of parallel transport, more or
less already used in some physical theories. The parallel transport concept
appropriately unifies the two above mentioned features:  {\it change} and a
suitable {\it projection}. In some cases this concept may be given a physical
interpretation of {\it conservation (balance) equation} (mainly
energy-momentum balance). The examples presented show how it has been used
and how it could be used as a field equations generating tool.  An important
feature, that deserves to be noted even at this moment, is that the
corresponding equations may become {\it nonlinear} in a natural way, so we
might be fortunately surprised by appearing of spatially finite (or
soliton-like) solutions (we shall recall such examples).

\section{The general Rule}
We begin with the algebraic structure to be used further in the bundle
picture. The basic concepts used are the {\it tensor product} $\otimes$ of
two linear spaces (we shall use the same term {\it linear space} for a
vector space over a field, and for a module over a ring, and from the context
it will be clear which case is considered) and {\it bilinear maps}. Let
$(U_1,V_1)$, $(U_2,V_2)$ and $(U_3,V_3)$ be three couples of linear spaces.
Let $\Phi:  U_1\times U_2\rightarrow U_3$ and $\varphi: V_1\times
V_2\rightarrow V_3$ be two bilinear maps. Then we can form
the elements $(u_1\otimes v_1)\in U_1\otimes V_1$ and $(u_2\otimes v_2)\in
U_2\otimes V_2$, and apply the given bilinear maps as follows:
$(\Phi,\varphi)(u_1\otimes v_1,u_2\otimes v_2)=\Phi(u_1,u_2)\otimes
\varphi(v_1,v_2)$. The obtained element is in  $U_3\otimes V_3$.

We give now the corresponding bundle picture. Let $M$ be a smooth
n-dimensiomal real manifold. We assume that the following vector bundles over
$M$ are constructed:
 $\xi_i, \eta_i$, with standard fibers $U_i,V_i$ and sets of
sections $Sec(\xi_i), Sec(\eta_i), i=1,2,3$.

Assume the two bundle maps are given: $(\Phi,id_M):
\xi_1\times\xi_2\rightarrow\xi_3$ and $(\varphi,id_M):
\eta_1\times\eta_2\rightarrow\eta_3$. Then if $\sigma_1$ and $\sigma_2$ are
sections of $\xi_1$ and $\xi_2$ respectively, and $\tau_1$ and $\tau_2$ are
sections of $\eta_1$ and $\eta_2$ respectively, we can form an element of
$Sec(\xi_3\otimes\eta_3)$:
\begin{equation}
(\Phi,\varphi)(\sigma_1\otimes\tau_1,\sigma_2\otimes\tau_2)=           
\Phi(\sigma_1,\sigma_2)\otimes\varphi(\tau_1,\tau_2).
\end{equation}

Let now $\tilde\xi$ be a new vector bundle on $M$ and $\sigma_2\in
Sec(\xi_2)$ is obtained by the action of the differential operator
$D:Sec(\tilde\xi)\rightarrow Sec(\xi_2)$ on a section $\tilde\sigma$ of
$\tilde\xi$, so we can form the section (instead of $\sigma_1$ we write just
$\sigma$)
$\Phi(\sigma,D\tilde\sigma)\otimes\varphi(\tau_1,\tau_2)\in
Sec(\xi_3\otimes\eta_3)$. We give now the following
\vskip 0.4cm
\noindent
{\bf Definition}:
The section $\tilde\sigma$ will be called
$(\Phi,\varphi;D)$-{\it parallel} with respect to $\sigma$ if
\begin{equation}
(\Phi,\varphi;D)(\sigma\otimes \tau_1, \tilde\sigma\otimes \tau_2)=
(\Phi,\varphi)(\sigma\otimes \tau_1, D\tilde\sigma\otimes \tau_2)=
\Phi(\sigma,D\tilde\sigma)\otimes\varphi(\tau_1,\tau_2)=0.     
\end{equation}
\vskip 0.4cm
This relation (2) we call the {\bf GENERAL RULE (GR)}, the map $\Phi$
"projects" the "changes" $D\tilde\sigma$ of the section $\tilde\sigma$ on the
 section $\sigma$ ($\sigma$ may depend on $\tilde\sigma$), and $\varphi$
"works" usually on the (local) bases of the bundles where $\sigma$ and
$D\tilde\sigma$ take values.  As an example of a differential operator we
note the particular case when $\tilde\xi$ is the bundle of exterior $p$-forms
on $M$ with the available differential operator {\it exterior derivative}
$\mathbf{d}:  \Lambda^p(M)\rightarrow\Lambda^{p+1}(M)$.  In the case of
the physically important example of Lie algebra $\mathfrak{g}$-valued
differential forms, with "$\Phi=$ exterior product" and "$\varphi=$ Lie
bracket $[,]$", $\xi_1=\Lambda^p(M)=\tilde{\xi}$, $\xi_2=\Lambda^{p+1}(M)$,
$\eta_1=\eta_2=M\times\mathfrak{g}$, the {\bf GR} (2) looks as follows:
\[
(\wedge,[,];\mathbf{d})(\alpha^i\otimes E_i, \beta^j\otimes E_j)=
(\wedge,[,])(\alpha^i\otimes E_i, \mathbf{d}\beta^j\otimes E_j)=
\alpha^i\wedge\mathbf{d}\beta^j\otimes [E_i,E_j]=0,
\]
where $\{E_i\}$ is a basis of $\mathfrak{g}$, and a summation over the
repeated indexes is understood.  Further we are going to consider particular
cases of the {\bf (GR)} (2) with explicitly defined differential operators
whenever they participate in the definition of the section of interest.

\section{The General Rule in Action}
\subsection{Classical mechanics}
We begin studying the potential strength of the {\bf GR} in the frame of
classical mechanics.
\vskip 0.4cm
\noindent
1. {\bf Integral invariance relations}
\vskip 0.2cm
These relations have been introduced and studied from the point of view
of applications in mechanics by Lichnerowicz [3].

We specify the bundles over the real finite dimensional manifold
$M$ introduced in sec.2:

$\xi_1=TM;\  \xi_2=T^*(M);\  \
\eta_1=\eta_2=\xi_3=\eta_3=M\times\mathbb{R},\  \text{denote}\
Sec(M\times \mathbb{R})\equiv C^{\infty}(M)$

$\Phi$=substitution operator, denoted by\  $i(X), X\in Sec(TM)$;

$\varphi$=point-wise product of functions.

We denote by $1$ the function $f(x)=1, x\in M$. Consider the sections
\newline $X\otimes 1\in Sec(TM\otimes(M\times \mathbb{R}));\ \ \alpha\otimes
1\in Sec(T^*M\otimes(M\times\mathbb{R}))$.  Then the {\bf GR} leads to
\begin{equation}
(\Phi,\varphi)(X\otimes 1, \alpha\otimes 1)=i(X)\alpha\otimes 1      
=i(X)\alpha=0.
\end{equation}
We introduce now the differential operator $\mathbf{d}$: if $\alpha$ is an
exact 1-form, $\alpha=\mathbf{d}f$, so that $\tilde{\xi}=M\times\mathbb{R}$,
the relation (3) becomes
\[
i(X)\alpha=i(X)\mathbf{d}f=X(f)=0,
\]
i.e. the derivative of $f$ along the vector field $X$ is equal to zero. So,
we obtain the well known relation, defining the first integrals  $f$ of the
dynamical system determined by the vector field $X$.  In this sense $f$ may
be called $(\Phi,\varphi,\mathbf{d})$-{\it parallel} with respect to $X$,
where $\Phi$ and $\varphi$ are defined above. In [3] $\alpha$ is a $p$-form,
$\alpha\in Sec(\Lambda^p(T^*M))$, but this does not change the validity of
the above relation (3).
\vskip 0.3cm
\noindent
2.{\bf Absolute and relative integral invariants}
\vskip 0.2cm
These quantities have been introduced and studied in mechanics by Cartan
[4]. By definition, a $p$-form $\alpha$ is called an {\it absolute integral
invariant} of the vector field $X$ if $i(X)\alpha=0$ and
$i(X)\mathbf{d}\alpha=0$. And $\alpha$ is called a {\it relative integral
invariant} of the field $X$ if $i(X)\mathbf{d}\alpha=0$. So, in our
terminology (the same bundle picture as above), we can call the relative
integral invariants of $X$ $(\Phi,\varphi;\mathbf{d})$-{\it parallel} with
respect to $X$, and the absolute integral invarians of $X$ have additionally
$(\Phi,\varphi)$-{\it parallelism} with respect to $X$, with $(\Phi,\varphi)$
as defined above.  A special case is when $p=n$, and $\omega\in\Lambda^n(M)$
is a volume form on $M$.
\newpage
\noindent 3. {\bf Symplectic mechanics}
\vskip 0.2cm
Symplectic manifolds are even dimensional and have a
distinguished nondegenerate closed $2$-form $\omega$, $\mathbf{d}\omega=0$.
This structure may be defined in terms of the {\bf GR} in the following way.
Choose $\xi_1=\eta_1=\eta_2=M\times\mathbb{R}$, $\xi_2=\Lambda^2(T^*M)$, and
$\mathbf{d}$ as a differential operator. Consider now the section $1\in
Sec(M\times\mathbb{R})$ and the section $\omega\otimes 1\in
Sec(\Lambda^2(T^*M))\otimes Sec(M\times\mathbb{R})$, with $\omega$ -
nondegenerate.  The map $\Phi$ is the product $f.\omega$ and the map
$\varphi$ is the product of functions. So, we have
\[
(\Phi,\varphi;\mathbf{d})(1\otimes1, \omega\otimes 1)=
1.\mathbf{d}\omega\otimes 1=\mathbf{d}\omega=0.
\]
Hence, the relation $\mathbf{d}\omega=0$ is equivalent to the requirement
$\omega$ to be $(\Phi,\varphi;\mathbf{d})$-{\it parallel} with respect to the
section $1\in Sec(M\times\mathbb{R})$.

The hamiltonian vector fields $X$ are defined by the condition
$L_X\omega=\mathbf{d}i(X)\omega=0$.
If $\Phi=\varphi$ is the point-wise product of functions
we have
\[
(\Phi,\varphi;\mathbf{d})(1\otimes 1,i(X)\omega\otimes 1)=
(\Phi,\varphi)(1\otimes 1,\mathbf{d}i(X)(\omega)\otimes 1)=
L_X\omega\otimes 1=L_X\omega=0.
\]
In terms of the {\bf GR} we can say that $X$ is hamiltonian if $i(X)\omega$ is
$(\Phi,\varphi;\mathbf{d})$-{\it parallel}.

The induced Poisson structure $\{f,g\}$, is
given in terms of the {\bf GR} by setting $\Phi=\omega^{-1}$,
where $\omega^{-1}.\omega=id_{TM}$,
$\varphi$=point-wise product of functions,
and $1\in Sec(M\times\mathbb{R})$. We get
\[
(\Phi,\varphi)(\mathbf{d}f\otimes 1, \mathbf{d}g\otimes 1)=
\omega^{-1}(\mathbf{d}f,\mathbf{d}g)\otimes 1.
\]
A closed 1-form $\alpha,\ \mathbf{d}\alpha=0$, is a first integral of the
hamiltonian system $Z$, $\mathbf{d}i(Z)\omega=0$, if $i(Z)\alpha=0$. In terms
of the {\bf GR} we can say that the first integrals $\alpha$ are
$(i,\varphi)$-parallel with respect to $Z$:
$(i,\varphi)(Z\otimes 1,\alpha\otimes 1)=i(Z)\alpha\otimes 1=0$.
From $L_Z\omega=0$ it follows
$L_Z\omega^{-1}=0$. The Poisson bracket $(\alpha,\beta)$ of two first
integrals $\alpha$ and $\beta$ is equal to
$(-\mathbf{d}\omega^{-1}(\alpha,\beta))$ [5]. The well known property that the
Poison bracket of two first integrals of $Z$ is again a first integral of $Z$
may be formulated as: the function $\omega^{-1}(\alpha,\beta)$ is
$(i,\varphi;\mathbf{d})$-parallel with respect to $Z$,

\[
(i,\varphi;\mathbf{d})(Z\otimes 1, \omega^{-1}(\alpha,\beta)\otimes 1)=
i(Z)\mathbf{d}\omega^{-1}(\alpha,\beta)\otimes 1=0.
\]

\subsection{Frobenius integrability theorems and linear connections}
\vskip 0.4cm
\noindent
1.{\bf Frobenius integrability theorems}
\vskip 0.2cm
Let $\Delta=(X_1,\dots,X_r)$ be a differential system on $M$, i.e. the vector
fields $X_i, i=1,\dots,r$ define a locally stable submodule of $Sec(TM)$ and
at every point $p\in M$ the subspace $\Delta_p^r\subset T_p(M)$ has dimension
$r$. Then $\Delta^r$ is called integrable if $[X_i,X_j]\in \Delta^r,
i,j=1,\dots,r$. Denote by $\Delta^{n-r}_p\subset T_p(M)$ the complimentary
subspace: $\Delta_p^r\oplus\Delta^{n-r}_p=T_p(M)$, and let $\pi:
T_p(M)\rightarrow \Delta^{n-r}_p$ be the corresponding projection. So,
the corresponding Frobenius integrability condition means $\pi([X_i,X_j])=0,
i,j=1,\dots,r$.

In terms of the {\bf GR} we set $D(X_i)=\pi\circ L_{X_i}$, $\Phi$="product of
functions and vector fields",  and
$\varphi$ again the pruduct of functions. The integrability
condition now is
\[
\begin{split}
&(\Phi,\varphi;D(X_i))
(1\otimes 1, X_j\otimes 1)\\&=(\Phi,\varphi)
(1\otimes 1,\pi([X_i,X_j]\otimes 1))=1.\pi([X_i,X_j])\otimes 1.1
=0,
\quad i,j=1,\dots,r.
\end{split}
\]

In the dual formulation we have the Pfaff system
$\Delta^*_{n-r}$, generated by the linearly independent 1-forms
$(\alpha_1,\dots,\alpha_{n-r})$, such that
$\alpha_m(X_i)=0, i=1,\dots r; m=1,\dots n-r$.
Then $\Delta^*_{n-r}$ is
integrable if $\mathbf{d}\alpha\wedge \alpha_1\wedge\dots \wedge
\alpha_{n-r}=0, \alpha\in \Delta^*_{n-r}$. In terms of {\bf GR} we set
$\varphi$ the same as above, $\Phi=\wedge$ and $\mathbf{d}$ as differential
operator.
\[ (\Phi,\varphi;\mathbf{d})(\alpha_1\wedge\dots
\wedge\alpha_{n-r}\otimes 1, \alpha\otimes 1)= \mathbf{d}\alpha\wedge
\alpha_1\wedge\dots \wedge\alpha_{n-r}\otimes 1=0.
\]

\vskip 0.4cm
\noindent
2. {\bf Linear connections}
\vskip 0.2cm
The concept of a linear connection in a vector bundle has proved to be of
great importance in geometry and physics. In fact, it allows to differentiate
sections of vector bundles along vector fields, which is a basic operation in
differential geometry, and in theoretical physics the physical fields are
represented mainly by sections of vector bundles. We recall now how one comes
to it.

Let $f:\mathbb{R}^n\rightarrow\mathbb{R}$ be a differentiable function. Then
we can find its differential $\mathbf{d}f$. The map $f\rightarrow\mathbf{d}f$
is $\mathbb{R}$-linear: $\mathbf{d}(\kappa.f)=\kappa.\mathbf{d}f$,
$\kappa \in \mathbb{R}$, and it has the
derivative property $\mathbf{d}(f.g)=f\mathbf{d}g+g\mathbf{d}f$. These two
properties are characteristic ones, and they are carried to the bundle
situation as follows.

Let $\xi$ be a vector bundle over $M$. We always have the trivial bundle
$\xi_o=M\times\mathbb{R}$. Consider now $f\in C^{\infty}(M)$ as a section of
$\xi_o$. We note that $Sec(\xi_o)=C^{\infty}(M)$ is a module over itself, so
we can form $\mathbf{d}f$ with the above two characteristic  properties. The
new object $\mathbf{d}f$ lives in the space $\Lambda^1(M)$ of 1-forms on $M$,
so it defines a linear map $\mathbf{d}f: Sec(TM)\rightarrow Sec(\xi_o),
\mathbf{d}f(X)=X(f)$.  Hence, we have a map $\nabla$ from $Sec(\xi_o)$ to
the 1-forms with values in $Sec(\xi_o)$, and this map has the above two
characteristic properties.  We say that $\nabla$ defines a linear connection
in the vector bundle $\xi_o$.

In the general case the sections $Sec(\xi)$ of the vector bundle $\xi$ form a
module over $C^{\infty}(M)$. So, a linear connection $\nabla$ in $\xi$ is a
$\mathbb{R}$-linear map $\nabla: Sec(\xi)\rightarrow \Lambda^1(M,\xi)$. In
other words, $\nabla$ sends a section $\sigma\in Sec(\xi)$ to a 1-form
$\nabla \sigma$ valued in $Sec(\xi)$ in such a way, that
\begin{equation}
\nabla(k\,\sigma)=k\,\nabla(\sigma), \quad                     
\nabla(f\,\sigma)=df\otimes\sigma+f\,\nabla(\sigma),
\end{equation}
where $k\in \mathbb{R}$ and $f\in C^{\infty}(M)$. If $X\in Sec(TM)$ then we
have the composition $i(X)\circ\nabla$, so that
\[
i(X)\circ\nabla(f\,\sigma)=X(f)\,\sigma+f\,\nabla_X(\sigma),
\]
where $\nabla_X(\sigma)\in Sec(\xi)$.

In terms of the {\bf GR} we put $\xi_1=TM=\tilde\xi$ and
$\xi_2=\Lambda^1(M)\otimes\xi$,
and $\eta_1=\eta_2=\xi_o$. Also, $\Phi(X,\nabla\sigma)=\nabla_X \sigma$ and
$\varphi(f,g)=f.g$. Hence, we obtain
\begin{equation}
(\Phi,\varphi;\nabla)(X\otimes 1, \sigma\otimes 1)           
(\Phi,\varphi)(X\otimes 1,(\nabla \sigma)\otimes 1)
=\nabla_X \sigma\otimes 1=\nabla_X \sigma,
\end{equation}
and the section $\sigma$ is called $\nabla$-{\it parallel} with respect to
$X$ if $\nabla_X\sigma=0$.

\vskip 0.4cm
\noindent
3. {\bf Covariant exterior derivative}
\vskip 0.2cm
The space of $\xi$-valued $p$-forms $\Lambda^p(M,\xi)$ on $M$ is isomorphic
to $\Lambda^p(M)\otimes Sec(\xi)$. So, if $(\sigma_1,\dots,\sigma_r)$ is a
local basis of $Sec(\xi)$, every $\Psi\in \Lambda^p(M,\xi)$ is represented by
$\psi^i\otimes \sigma_i, i=1,\dots,r$, where $\psi^i\in \Lambda^p(M)$.
Clearly the space $\Lambda(M,\xi)=\Sigma^n_{p=0}\Lambda^p(M,\xi)$, where
$\Lambda^o(M,\xi)=Sec(\xi)$, is a
$\Lambda(M)=\Sigma^n_{p=0}\Lambda^p(M)$-module:
$\alpha.\Psi=\alpha\wedge\Psi=(\alpha\wedge\psi^i)\otimes \sigma_i$.

A linear connection $\nabla$ in $\xi$ generates covariant exterior
derivative $\mathbf{D}: \Lambda^p(M,\xi)\rightarrow\Lambda^{p+1}(M,\xi)$ in
$\Lambda(M,\xi)$ according to the rule
\[
\begin{split}
\mathbf{D}\Psi&=\mathbf{D}(\psi^i\otimes \sigma_i)=
\mathbf{d}\psi^i\otimes \sigma_i+(-1)^p \psi^i\wedge\nabla(\sigma_i)\\
&=(\mathbf{d}\psi^i+(-1)^p \psi^j\wedge\Gamma_{\mu j}^i dx^\mu)\otimes\sigma_i
=(\mathbf{D}\Psi)^i\otimes\sigma_i.
\end{split}
\]
We may call now a $\xi$-valued $p$-form $\Psi$ $\nabla$-{\it parallel} if
$\mathbf{D}\Psi=0$, and $(X,\nabla)$-{\it parallel} if
$i(X)\mathbf{D}\Psi=0$. This definition extends in a natural way to
$q$-vectors with $q\le p$. Actually, the substitution operator $i(X)$ extends
to (decomposable) $q$-vectors $X_1\wedge X_2\wedge\dots\wedge X_q$ as
follows:
\[
i(X_1\wedge X_2\wedge\dots\wedge X_q)\Psi =i(X_q)\circ
i(X)_{q-1}\circ\dots\circ i(X_1)\Psi,
\]
and extends to nondecomposable
$q$-vectors by linearity. Hence, if $\Theta$ is a section of
$\Lambda^q(TM)$ we may call $\Psi$ $(\Theta,\nabla)$-{\it parallel} if
$i(\Theta)\mathbf{D}\Psi=0$.

Denote now by $L_\xi$ the vector bundle of (linear) homomorphisms $(\Pi,id):
\xi\rightarrow \xi$, and let $\Pi\in Sec(L_\xi)$. Let
$\chi \in Sec(\Lambda^q(TM)\otimes L_\xi)$ be represented as
$\Theta\otimes\Pi$. The map $\Phi$ will act as: $\Phi(\Theta,\Psi)=
i(\Theta)\Psi$, and the map $\varphi$ will act as: $\varphi(\Pi,\sigma_i)=
\Pi(\sigma_i)$.
So, if $\nabla(\sigma_k)=\Gamma^j_{\mu k}dx^\mu\otimes
\sigma_j$, we may call $\Psi$ $(\nabla)$-{\it parallel} with respect to
$\chi$ if
\begin{equation}
(\Phi,\varphi;\mathbf{D})(\Theta\otimes\Pi,\Psi=\psi^i\otimes\sigma_i)=
(\Phi,\varphi)(\Theta\otimes\Pi,(\mathbf{D}\Psi)^i\otimes\sigma_i)=     
i(\Theta)(\mathbf{D}\Psi)^i\otimes\Pi(\sigma_i)=0.
\end{equation}
If we have isomorphisms $\otimes^p TM\backsim
\otimes^p T^*M, p=1,2,\dots$, defined in some natural way (e.g. through a
metric tensor field), then to any $p$-form $\alpha$ corresponds unique
$p$-vector $\tilde\alpha$. In this case we may talk about "$\backsim$"-
{\it autopaparallel} objects with respect a (point-wise) bilinear map
$\varphi:  (\xi\times\xi)\rightarrow \eta$, where $\eta$ is also a vector
bundle over $M$. So, $\Psi=\alpha^k\otimes\sigma_k\in \Lambda^p(M,\xi)$ may
be called $(i,\varphi;\nabla)$-{\it autoparallel} with respect to the
isomorphism "$\backsim$" if
\begin{equation}
\begin{split}
&(i,\varphi;\nabla)
(\tilde\alpha^k\otimes \sigma_k,\alpha^m\otimes\sigma_m)\\
&=i(\tilde\alpha^k)\mathbf{d}\alpha^m\otimes\varphi(\sigma_k,\sigma_m)+
(-1)^p i(\tilde\alpha^k)(\alpha^j\wedge \Gamma^m_{\mu j}dx^\mu)
\otimes\varphi(\sigma_k,\sigma_m)\\                                   
&=\big[i(\tilde\alpha^k)\mathbf{d}\alpha^m+
(-1)^p i(\tilde\alpha^k)(\alpha^j\wedge \Gamma^m_{\mu j}dx^\mu)\big]
\otimes\varphi(\sigma_k,\sigma_m)=0.
\end{split}
\end{equation}
\noindent
Although the above examples do not, of course, give a complete list of the
possible applications of the {\bf GR} (2), they will serve as a good basis
for the physical applications we are going to consider further.

\section{Physical applications of GR}

{\bf 1. Autoparallel vector fields and 1-forms}
\vskip 0.2cm
In nonrelativistic and relativistic mechanics the vector fields $X$ on a
manifold $M$ are the local representatives (velocity vectors) of the
evolution trajectories for point-like objects.  The condition that a particle
is {\it free} is mathematically represented by the requirement that the
corresponding vector field $X$ is autoparallel with respect to a given
connection $\nabla$ (covariant derivative) in $TM$:
\begin{equation}
i(X)\nabla X=0,\quad
\text{or in components},\quad
X^\sigma \nabla_\sigma X^\mu +\Gamma^\mu_{\sigma\nu}X^\sigma X^\nu=0.   
\end{equation}
In view of the physical interpretation of $X$ as velocity vector field the
usual latter used instead of $X$ is $u$. The above equation (8) presents a
system of nonlinear partial differential equations for the components
$X^\mu$, or $u^\mu$. When reduced to 1-dimensional submanifold which is
parametrised locally by the appropriately chosen parameter $s$, (8) gives a
system of ordinary differential equations:
\begin{equation}
\frac{d^2 x^\mu}{ds^2}+\Gamma^\mu_{\sigma\nu}\frac{dx^\nu}{ds}       
\frac{dx^\nu}{ds}=0,
\end{equation}
and (9) are known as ODE defining the geodesic (with respect to $\Gamma$)
lines in $M$. When $M$ is reimannian with metric tensor $g$ and $\Gamma$ the
corresponding Levi-Civita connection, i.e. $\nabla g=0$ and
$\Gamma^\mu_{\nu\sigma}=\Gamma^\mu_{\sigma\nu}$, then the solutions of (9)
give the extreme (shortest or longest) distance $\int^b_a ds$
between the two points $a,b\in M$, so (9) are equivalent to
\[
\delta\left(\int^b_a ds\right)=
\delta\left(\int^b_a
\sqrt{g_{\mu\nu}\frac{dx^\mu}{ds}\frac{dx^\nu}{ds}}\right)=0.
\]
A system of particles that move along the solutions to (9) with $g$-the
Minkowski metric and $g_{\mu\nu}\frac{dx^\mu}{ds}\frac{dx^\nu}{ds}>0$,
is said to form an {\it inertial frame of reference}.

It is interesting to note that the system (8) has (3+1)-soliton-like
(even spatially finite) solutions on Minkowski space-time [6]. In fact, in
canonical coordinates $(x^1,x^2,x^3,x^4)=(x,y,z,\xi=ct)$ let
$u^\mu=(0,0,\pm \frac vc f,f)$
be the components of $u$, where $0<v=const<c$, and $c$ is the velocity
of light, so $\frac vc < 1$ and $u^\sigma u_\sigma =
\left(1-\frac{v^2}{c^2}\right)f^2>0$. Then every function $f$ of the kind
\[
f(x,y,z,\xi)=f\left(x,y,\alpha(z\mp\frac vc
\xi)\right),\  \alpha=const, \quad\text{for example}\quad
\alpha=\frac{1}{\sqrt{1-\frac{v^2}{c^2}}},
\]
defines a slution to (8).  If $u_\sigma u^\sigma =0$ then equations (8)
($X=u$), are equivalent to $u^\mu(\mathbf{d}u)_{\mu\nu}=0$, where
$\mathbf{d}$ is the exterior derivative. In fact, since the connection used
is riemannian, we have $0=\nabla_\mu\frac12(u^\nu u_\nu)=u^\nu\nabla_\mu
u_\nu$, so the relation $u^\nu\nabla_\nu u_\mu -u^\nu\nabla_\mu u_\nu=0$
holds and is obviously equal to $u^\mu(\mathbf{d}u)_{\mu\nu}=0$. The
soliton-like solution is defined by $u=(0,0,\pm f,f)$ where the function $f$
is of the form
\[
f(x,y,z,\xi)=f(x,y,z\mp \xi).
\]
Clearly, for every autoparallel vector field $u$ (or one-form $u$) there
exists a canonical coordinate system on the Minkowski space-time, in
which $u$ takes such a simple form: $u^\mu=(0,0,\alpha f,f), \alpha=const$.
The dependence of $f$ on the three spatial coordinates $(x,y,z)$ is arbitrary
, so it is allowd to be chosen {\it soliton-like} and, even, {\it finite}.
Let now $\rho$ be the mass-energy density function, so that
$\nabla_\sigma(\rho u^\sigma)=0$ gives the mass-energy conservation, i.e. the
function $\rho$ defines those properties of our physical system which
identify the system during its evolution. In this way the tensor conservation
law
\[
\nabla_\sigma(\rho u^\sigma u^\mu)=(\nabla_\sigma \rho u^\sigma)u^\mu+
\rho u^\sigma\nabla_\sigma u^\mu=0
\]
describes the two aspects of the physical system: its dynamics through
equations (8) and its mass-energy conservation properties.

The properties described give a connection between free point-like objects
and (3+1) soliton-like autoparallel vector fields on Minkowski space-time.
Moreover, they suggest that extended free objects with more complicated
space-time dynamical structure may be described by some appropriately
generalized concept of autoparallel mathematical objects.
\vskip 0.4cm
{\bf 2. Electrodynamics}
\vskip 0.2cm
{\bf 2.1 Maxwell equations}

The Maxwell equations $\mathbf{d}F=0, \mathbf{d}*F=0$ in their 4-dimensional
formulation on Minkowski space-time $(M,\eta), sign(\eta)=(-,-,-,+)$ and the
Hodge $*$ is defined by $\eta$, make use of the exterior derivative as a
differential operator.  The field has, in general, 2 components
$(F,*F)$, so the interesting bundle is $\Lambda^2(M)\otimes V$, where $V$ is
a real 2-dimensional vector space. Hence the adequate mathematical field
will look like $\Omega=F\otimes e_1+*F\otimes e_2$ [7], where $(e_1,e_2)$ is a
basis of $V$. The exterior derivative acts on $\Omega$ as:
$\mathbf{d}\Omega=\mathbf{d}F\otimes e_1+\mathbf{d}*F\otimes e_2$, and the
equation $\mathbf{d}\Omega=0$ gives the vacuum Maxwell equations.

In order to interpret in terms of the above given general view ({\bf GR})
on parallel objects with respect to given sections of vector bundles and
differential operators we consider the sections (see the above introdused
notation) $(1\times1, \Omega\times 1)$ and the differential operator
$\mathbf{d}$. Hence, the {\bf GR} acts as follows:
\[
(\Phi,\varphi;\mathbf{d})(1\times1, \Omega\times 1)=
(\Phi,\varphi)(1\times1,\mathbf{d}\Omega\times 1)=
(1.\mathbf{d}\Omega\otimes 1.1)
\]
The corresponding $(\Phi,\varphi;\mathbf{d})$-parallelism leads to
$\mathbf{d}\Omega=0$. In presence of electric $\mathbf{j}$ and magnetic
$\mathbf{m}$
currents, considered as 3-forms, the parallelism condition does not hold and
on the right-hand side we'll have non-zero term, so the full condition is
\begin{equation}
(\Phi,\varphi)(1\times1, (\mathbf{d}F\otimes e_1+
\mathbf{d}*F\otimes e_2)\times 1)=
(\Phi,\varphi;\mathbf{d})(1\times1, (\mathbf{m}\otimes e_1+        
\mathbf{j}\otimes e_2)\times 1)
\end{equation}
The case $\mathbf{m}=0, F=\mathbf{d}A$ is, obviously a special case.

\vskip 0.2cm
{\bf 2.2 Extended Maxwell equations}

The extended Maxwell equations (on Minkowski space-time) in vacuum read [8]:
\begin{equation}
F\wedge *\mathbf{d}F=0,\quad (*F)\wedge(*\mathbf{d}*F)=0,\quad      
F\wedge(*\mathbf{d}*F)+(*F)\wedge(*\mathbf{d}F)=0
\end {equation}
They may be expressed through the {\bf GR} in the following way.
On $(M,\eta)$ we have the bijection between $\Lambda^2(TM)$ and
$\Lambda^2(T^*M)$ defined by $\eta$, which we denote by
$\tilde F\leftrightarrow F$. So, equations (11) are equivalent to
\[
i(\tilde F)\mathbf{d}F=0,\quad i(\widetilde{*F})\mathbf{d}*F=0,\quad
i(\tilde F)\mathbf{d}*F+i(\widetilde{*F})\mathbf{d}F=0.
\]
We consider the
sections $\tilde\Omega=\tilde F\otimes e_1+\widetilde{*F}\otimes e_2$ and
$\Omega=F\otimes e_1+*F\otimes e_2$ with the differential operator
$\mathbf{d}$. The maps $\Phi$ and $\varphi$ are defined as:
$\Phi$ is the substitution operator $i$, and $\varphi=\vee$ is
the symmetrized tensor product in $V$. So we obtain
\begin{equation}
\begin{split} &(\Phi,\varphi;\mathbf{d})(\tilde F\otimes
e_1+\widetilde{*F}\otimes e_2, F\otimes e_1+*F\otimes e_2)\\ &=i(\tilde
F)\mathbf{d}F\otimes e_1\vee e_1+ i(\widetilde{*F})\mathbf{d}*F\otimes  
e_2\vee e_2+ (i(\tilde F)\mathbf{d}*F+i(\widetilde{*F})\mathbf{d}F)
\otimes e_1\vee e_2=0.
\end{split}
\end{equation}
Equations (12) may be written down also as
$\big((i,\vee)\tilde\Omega\big)\mathbf{d}\Omega=0$.

Equations (12) are physically interpreted as describing locally the intrinsic
energy-momentum exchange between the two components $F$ and $*F$ of $\Omega$:
the first two equations
$i(\tilde F)\mathbf{d}F=0$ and $i(\widetilde *F)\mathbf{d}*F=0$
say that every component keeps locally its
energy-momentum, and the third equation $i(\tilde F)\mathbf{d}*F+i(\widetilde
*F)\mathbf{d}F=0$ says (in accordance with the first two) that if $F$
transfers energy-momentum to $*F$, then $*F$ transfers the same quantity
energy-momentum to $F$.

If the field exchanges (loses or gains) energy-momentum with some external
systems, Extended Electrodynamics describes the potential abilities of the
external systems to gain or lose energy-momentum from the field by means of 4
one-forms (currents) $J_a, a=1,2,3,4$, and explicitly the exchange is
given by [8]
\begin{equation}
i(\tilde F)\mathbf{d}F=i(\tilde J_1)F,\ \
i(\widetilde {*F})\mathbf{d}*F=i(\tilde J_2)F,\ \
i(\tilde F)\mathbf{d}*F+i(\widetilde {*F})\mathbf{d}F=             
i(\tilde J_3)F+i(\tilde J_4)*F.
\end{equation}
It is additionally assumed that every couple $(J_a,J_b)$ defines
a completely integrable Pfaff system, i.e. the following equations hold:
\begin{equation}
\mathbf{d}J_a\wedge J_a\wedge J_b=0,\ \ a,b=1,\dots,4.             
\end{equation}
The system (12) has (3+1)-localised photon-like (massless) solutions [7], and
the system (13)-(14) admits a larage family of (3+1)-soliton solutions [8].

\vskip 0.4cm
{\bf 3. Yang-Mills theory}
\vskip 0.2cm
{\bf 3.1 Yang-Mills equations}

In this case the field is a connection, represented locally by its connection
form $\omega\in \Lambda^1(M)\otimes\mathfrak{g}$, where $\mathfrak{g}$ is the
Lie algebra of the corresponding Lie group $G$.
If $\mathbf{D}$ is the corresponding
covariant derivative, and $\Omega=\mathbf{D}\omega$ is the curvature ,
then Yang-Mills equations read
$\mathbf{D}*\Omega=0$. The formal difference with the Maxwell case
is that $G$ may NOT be commutative, and may have, in general, arbitrary
finite dimension. So, the two sections are $1\otimes 1$ and
$*\Omega\otimes 1$, the maps $\Phi$ and $\varphi$ are product of functions
and the differential operator is $\mathbf{D}$. So, we may write
\begin{equation}
(\Phi,\varphi;\mathbf{D})(1\otimes 1, *\Omega\otimes 1)=      
\mathbf{D}*\Omega\otimes 1=0.
\end{equation}
Of course, equations (13) are always coupled to the Bianchi identity
$\mathbf{D}\Omega=0$.

\vskip 0.3cm
{\bf 3.2 Extended Yang-Mills equations}
\vskip 0.2cm
The extended Ynag-Mills equations are written down in analogy
with the extended Maxwell equations.
The field of interest is an arbitrary 2-form  $\Psi$ on $(M,\eta)$ with
values in a Lie algebra $\mathfrak{g}$, $\dim(\mathfrak{g})=r$. If
$\{E_i\}, i=1,2,\dots,r$ is a basis of $\mathfrak{g}$ we have
$\Psi=\psi^i\otimes E_i$ and $\tilde\Psi=\tilde\psi^i\otimes E_i$. The map
$\Phi$ is the substitution operator, the map $\varphi$ is the corresponding
Lie product $[,]$, and the differential operator is the exterior covariant
derivative with respect to a given connection $\omega$:
$\mathbf{D}\Psi=\mathbf{d}\Psi+[\omega,\Psi]$. We obtain
\begin{equation}
(\Phi,\varphi;\mathbf{D})(\tilde\psi^i\otimes E_i,\psi^j\otimes E_j)=
i(\tilde\psi^i)(\mathbf{d}\psi^m+\omega^j\wedge\psi^k\,C_{jk}^m)      
\otimes[E_m,E_i]=0,
\end{equation}
where $C_{jk}^m$ are the corresponding structure constants.
If the connection is the trivial one, then $\omega=0$ and
$\mathbf{D}\rightarrow \mathbf{d}$, so, this equation reduces to
\begin{equation}
i(\tilde\psi^i)\mathbf{d}\psi^j\,C_{ij}^k\otimes E_k=0         
\end{equation}
If, in addition, instead of $[,]$ we assume for $\varphi$ some bilinear map
$f:\mathfrak{g}\times\mathfrak{g}\rightarrow\mathfrak{g}$, such that in the
basis $\{E_i\}$ $f$ is given by $f(E_i,E_i)=E_i$, and $f(E_i,E_j)=0$ for
$i\neq j$ the last relation reads
\begin{equation}
i(\tilde\psi^i)\mathbf{d}\psi^i\otimes E_i=0,\quad i=1,2,\dots,r.  
\end{equation}
The last equations (18) define the components $\psi^i$ as independent 2-forms
(of course $\psi^i$ may be arbitrary $p$-forms). If the bilinear map
$\varphi$ is chosen to be the symmetrized tensor product
$\vee:\mathfrak{g}\times \mathfrak{g}\rightarrow \mathfrak{g}\vee
\mathfrak{g}$, we obtain
\begin{equation}
i(\psi^i)\mathbf{d}\psi^j\otimes E_i\vee E_j=0,\quad i\leqq j=1,\dots,r. 
\end{equation}
Equations (17) and (19) may be used to model bilinear interaction among the
components of $\Psi$. If the terms $i(\psi^i)\mathbf{d}\psi^j\otimes E_i\vee
E_j$ have the physical sense of energy-momentum exchange we may say that
every component $\psi^i$ gets locally as much energy-momentum from $\psi^j$
as it gives to it. Since $C^k_{ij}=-C^k_{ji}$,
equations (17) consider only the case $i<j$,
while equations (19) consider $i\leq j$, in fact, for every
$i,j=1,2,\dots,r$ we obtain from (19)
\[
i(\tilde\psi^i)\mathbf{d}\psi^i=0,\ \ \text{and}\ \
i(\tilde\psi^i)\mathbf{d}\psi^j+i(\tilde\psi^j)\mathbf{d}\psi^i=0.
\]
Clearly, these last equations may be considered as a natural generalization of
equations (12), so spatial soliton-like solutions are expectable.

\vskip 0.4cm
{\bf 4. General Relativity}
\vskip 0.2cm
In General Relativity the field function of interest is in a definite
sense identified with a pseudometric $g$ on a 4-dimensional manifold,
and only those $g$ are considered as appropriate to describe the real
gravitaional fields which satisfy the equations
$R_{\mu\nu}=0$, where $R_{\mu\nu}$ are the components of the Ricci tensor.
The main mathematical object which detects possible gravity is the Riemann
curvature tensor $R_{\alpha\mu,\beta\nu}$, which is a second order nonlinear
differential operator $R:g\rightarrow R(g)$. We define the map $\Phi$ to be
the contraction, or taking a {\it trace}:
$$
\Phi:(g_{\alpha\beta},R_{\alpha\mu,\beta\nu})
=g^{\alpha\beta}R_{\alpha\mu,\beta\nu}=R_{\mu\nu},
$$
so it is obviouly bilinear. The map $\varphi$ is a product of functions, so
the {\bf GR} gives
\begin{equation}
(\Phi,\varphi;R)(g\otimes 1,g\otimes 1)=\Phi(g,R(g))\otimes 1    
=Ric(R(g))\otimes 1=0.
\end{equation}

\vskip 0.4cm
{\bf 5. Schr\"odinger equation}
\vskip 0.2cm
The object of interest in this case is a map $\Psi: \mathbb{R}^4\rightarrow
\mathbb{C}$, and $\mathbb{R}^4=\mathbb{R}^3\times\mathbb{R}$ is
parametrized by the canonical coordinates $(x,y,z;t)$, where $t$ is the
(absolute) time "coordinate". The operator $D$ used here is
$$
D=i\hbar\frac{\partial }{\partial t}-\mathbf{H},
$$
where $\mathbf{H}$ is the corresponding {\it hamiltonian}. The maps $\Phi$
and $\varphi$ are products of functions, so the {\bf GR} gives
\begin{equation}
(\Phi,\varphi;\mathbf{D})(1\otimes 1,\Psi\otimes 1)=
\left(1\otimes\left(i\hbar\frac{\partial \Psi}                  
{\partial t}-\mathbf{H}\Psi\right)\right)\otimes 1=0.
\end{equation}

\vskip 0.4cm
{\bf 6. Dirac equation}
\vskip 0.2cm
The original free Dirac
equation on the Minkowski space-time $(M,\eta)$ makes use of the following
objects:
$\mathbb{C}^4$ - the canonical 4-dimensional complex vector space,
$L_{\mathbb{C}^4}$-the space of $\mathbb{C}$-linear maps
$\mathbb{C}^4\rightarrow\mathbb{C}^4$, $\Psi\in Sec(M\times \mathbb{C}^4)$,
$\gamma\in Sec(T^*M\otimes L_{\mathbb{C}^4})$, and the usual differential
$\mathbf{d}: \psi^i\otimes e_i\rightarrow \mathbf{d}\psi^i\otimes e_i$, where
$\{e_i\}, i=1,2,3,4$, is a basis of $\mathbb{C}^4$. We identify further
$L_{\mathbb{C}^4}$ with $(\mathbb{C}^4)^*\otimes \mathbb{C}^4$ and if
$\{\varepsilon^i\}$ is a basis of $(\mathbb{C}^4)^*$, dual to $\{e_i\}$, we
have the basis $\varepsilon^i\otimes e_j$ of $L_{\mathbb{C}^4}$. Hence, we
may write
\[
\gamma=\gamma_{\mu i}^j dx^\mu\otimes(\varepsilon^i\otimes e_j),
\]
and
\[
\begin{split}
\gamma(\Psi)&=
\gamma_{\mu i}^j dx^\mu\otimes(\varepsilon^i\otimes e_j)(\psi^k\otimes e_k)\\
&=\gamma_{\mu i}^j dx^\mu\otimes \psi^k<\varepsilon^i,e_k>e_j=
\gamma_{\mu i}^j dx^\mu\otimes \psi^k\delta^i_k e_j=
\gamma_{\mu i}^j \psi^i dx^\mu\otimes e_j.
\end{split}
\]
The 4 matrices $\gamma_\mu$ satisfy
$\gamma_\mu\gamma_\nu+\gamma_\nu\gamma_\mu=\eta_{\mu\nu}id_{\mathbb{C}^4}$,
so they are nondegenerate: $det(\gamma_\mu)\neq 0, \mu=1,2,3,4$,
and we can find $(\gamma_\mu)^{-1}$ and introduce $\gamma^{-1}$ by
\[
\gamma^{-1}=((\gamma_\mu)^{-1})_i^jdx^\mu\otimes(\varepsilon^i\otimes e_j)
\]
We introduce now the differential operators
$\mathcal{D}^{\pm}:Sec(M\times\mathbb{C}^4)\rightarrow Sec(T^*M\otimes
\mathbb{C}^4)$ through the formula:
$\mathcal{D}^{\pm}=i\mathbf{d}\pm\frac 12 m\gamma^{-1}, i=\sqrt{-1},
m\in \mathbb{R}$.
The corresponding maps are: $\Phi=\eta$, $\varphi:
L_{\mathbb{C}^4}\times\mathbb{C}^4\rightarrow\mathbb{C}^4$ given by
$\varphi(\alpha^*\otimes\beta,\rho)=<\alpha^*,\rho>\beta$.  We obtain

\begin{equation}
\begin{split}
&(\Phi,\varphi;\mathcal{D}^{\pm})(\gamma,\Psi)=
(\Phi,\varphi)(\gamma^j_{\mu i}dx^\mu\otimes(\varepsilon^i\otimes e_j),
i\frac{\partial \psi^k}{\partial x^\nu}dx^\nu\otimes e_k\pm\frac 12 m
(\gamma_\nu^{-1})^s_r dx^\nu\otimes(\varepsilon^r\otimes e_s)\psi^m e_m)\\
&=i\gamma^j_{\mu i}\frac{\partial \psi^k}{\partial x^\nu}\eta(dx^\mu,dx^\nu)
<\varepsilon^i,e_k>e_j\pm\frac 12 m\gamma^j_{\mu i}(\gamma_\nu^{-1})^s_r
\psi^r\eta(dx^\mu,dx^\nu)<\varepsilon^i,e_s>e_j\\
&=i\eta^{\mu\nu}\gamma^j_{\mu i}\frac{\partial \psi^k}
{\partial x^\nu}\delta^i_k e_j\pm                                        
\frac 12 m\eta^{\mu\nu}\gamma^j_{\mu i}(\gamma_\nu^{-1})^s_r\psi^r
\delta^i_s e_j\\
&=i\gamma^{\mu j}_i\frac{\partial \psi^i}{\partial x^\mu} e_j\pm
\frac 12 m(-2\delta^j_r\psi^r)e_j
=\left(i\gamma^{\mu j}_i \frac{\partial \psi^i}{\partial
x^\mu}\mp m\psi^j\right)e_j=0
\end{split}
\end{equation}
In terms of parallelism we can say that the Dirac equation is equavalent to
the requirement the section $\Psi\in Sec(M\times \mathbb{C}^4)$ to be
($\eta,\varphi;\mathcal{D}^{\pm}$)-parallel with respect to the given
$\gamma\in Sec(M\times L_{\mathbb{C}^4})$. Finally, in presence of external
electromagnetic field $\mathbf{A}=A_\mu dx^\mu$ the differential operators
$\mathcal{D}^{\pm}$ modify to
$\mathfrak{D}^{\pm}=i\mathbf{d}-
e\mathbf{A}\otimes id_{\mathbb{C}^4}\pm \frac 12 m \gamma^{-1}$,
where $e$ is the electron charge.

\section{Conclusion}
It was shown that the {\bf GR}, defined by relation (2), naturally
generalizes the geometrical concept of parallel transport,  and that it may
be successfully used as a unified tool to represent formally important
equations in theoretical physics. If $\Psi$ is the object of interest then
the {\bf GR} specifies mainly the following things: the change
$\mathbf{D}\Psi$ of $\Psi$, the object $\Psi_1$ with respect to which we
consider the change, the projection of the change $\mathbf{D}\Psi$ on
$\Psi_1$, and the bilinear amp $\varphi$ determines the space where
the final object lives.  When $\Psi=\Psi_1$ we may speak about {\it
autoparallel} objects, and in this case, as well as when the differential
operator $\mathbf{D}$ depends on $\Psi$ and its derivatives, we obtain {\it
nonlinear} equation(s).  In most of the examples considered the main
differential operator used was the usual differential {\bf d} and its
covariant generalization.

In the case of vector fields and one-forms on the Minkowski space-time we
recalled our previous result that among the corresponding autoparallel vector
fields there are finite (3+1) soliton-like ones, time-like, as well as
isotropic.  This is due to the fact that the trajectories of these fields
define straight lines, so their "transverse" components should be zero if
the spatially finite (or localized) configuration must move along these
trajectories.  Moreover, the corresponding equation can be easily modified to
be interpreted as local energy-momentum conservation relation.

It was further shown that Maxwell vacuum equations appear as
$\mathbf{d}$-parallel, i.e. {\it without specifying any projection
procedure}.  This determines their {\it linear} nature and leads to the lack
of spatial soliton-like solutions. The vacuum extended Maxwell equations (11)
are naturally cast in the form of autoparallel (nonlinear) equations, and, as
it was shown in our former works, they admit photon-like (3+1) {\it spatially
finite} and {\it spatially localized} solutions, and some of them admit
naturally defined spin properties [9]. The general extended Maxwell equations
(13)-(14) may also be given such a form if we replace the operator
$\mathbf{d}$ with $\mathbf{d}-i(J_a)$ in (13), and (3+1) soliton solutions of
this system were also found [8].

The Yang-Mills equations were also described in this way. The introduced in
this paper extended Yang-Mills equations (16)-(19) are expected to give
spatial soliton solutions. The vacuum Einstein equations of General
Relativity also admit such a formulation.

In quantum physics the Schr\"odinger equation admits the "parallel"
formulation without any projection. A bit more complicated was to put the
Dirac equation in  this formulation, and this is due to the bit more
complicated mathematical structure of $\gamma=\gamma_{\mu i}^j
dx^\mu\otimes(\varepsilon^i\otimes e_j)$.

These important examples make us think that the introduced in this paper
extended concept for $(\Phi,\varphi;\mathbf{D})$-{\it parellel objects} as a
natural generalization of the existing geometrical concept for
$\nabla$-parallel objects, may be successfully used in various directions, in
particulr, in searching for appropriate nonlinearizations of the existing
linear equations in theoretical and mathematical physics. It may also turn
out to find it useful in looking for appropriate lagrangeans in some cases.
In our view, as we pointed out in the Introduction, this is due to the fact
that it expresses in a unified manner the dual {\it change-conservation}
nature of the physical objects.

\vskip 1.5cm
{\bf References}
\vskip 0.6cm
[1]. {\bf Kol$\acute{\mathrm{a}}\check{\mathrm{r}}$, I., Michor, P.,
Slov$\acute{\mathrm{a}}$k, J}.,  {\it Natural Operations in Differential
Geometry}, Springer-Verlag, 1993

[2]. {\bf Greub, W., Halperin, S., Vanstone, R}., {\it Connections, Curvature
and Cohomology II}, Academic Press, New York and London, 1973

[3]. {\bf Lichn$\acute{\mathrm{e}}$rowicz, A}., {\it Les Relations
Int\'egrales d'Invariance et leurs Applications $\grave{\mathrm{a}}$ la
Dinamique}, Bull. Sc. Math., 70 (1946), 82-95

[4]. {\bf Cartan, E}., {\it Le\c{c}ons sur les Invariants int\'egraux},
Hermann \& Fils, Paris, 1922

[5]. {\bf Godbillon, C.}, {\it G\'eom\'etrie Differentielle et M\'echanique
Analytique}, Hermann, Paris, 1969

[6]. {\bf Donev, S}., {\it Autoclosed Differential Forms and (3+1)-Solitary
Waves}, Bulg. J. Phys., 15 (1988), 419-426

[7]. {\bf Donev, S., Tashkova, M}., {\it Energy-momentum Directed
Nonlinearization of Maxwell's Pure Field Equations}, Proc. R. Soc. Lond. A,
443 (1993), 301-312

[8]. {\bf Donev, S., Tashkova, M}., {\it Energy-momentum Directed
Nonlinearization of Maxwell's Equations in the Case of a Continuous Madia},
Proc. R. Soc.  Lond.  A, 443 (1995), 281-291

[9]. {\bf Donev, S}., {\it Screw Photon-like (3+1)-Solitons in Extended
Electrodynamics}, LANL e-print, hep-th/0104088 (a short version is accepted
for publ. in EPJ)

\end{document}